\documentclass[aps,preprint,showpacs,preprintnumbers,amsmath,amssymb,nofootinbib]{revtex4}
\usepackage{xcolor}
\usepackage{epsf}
\usepackage{epsfig}
\usepackage{graphicx}
\usepackage{latexsym}
\usepackage{bm}
\usepackage{amssymb}

\begin{document}


\preprint{to be submitted to PRA}

\title {Transfer ionization and its sensitivity to the ground-state wave function}

\author{M.~S.~Sch\"{o}ffler$^{1}$}
\email{schoeffler@atom.uni-frankfurt.de}
\author{O. Chuluunbaatar$^{2,3}$}
\author{Yu. V. Popov$^{4}$}
\author{S. Houamer$^{5}$}
\author{J.~Titze$^{1}$}
\author{T.~Jahnke$^{1}$}
\author{L. Ph. H.~Schmidt$^{1}$}
\author{O.~Jagutzki$^{1}$}
\author{A. G. Galstyan$^{6}$}
\author{A. A. Gusev$^{2}$}

\affiliation{$^1$ Institut f\"ur Kernphysik, Universit\"at
Frankfurt, Max-von-Laue-Str. 1, 60438 Frankfurt, Germany}

\affiliation{$^2$ Laboratory of Information Technologies, Joint
Institute for Nuclear Research, Dubna, Moscow region 141980,
Russia}

\affiliation{$^3$ School of Mathematics and Computer Sciences,
National University of Mongolia, UlaanBaatar, Mongolia}

\affiliation{$^4$ Skobeltsyn Institute of Nuclear Physics,
Lomonosov Moscow State University, Moscow 119991, Russia}

\affiliation{$^5$ Laboratoire de Physique Quantique et Syst\'emes
Dynamiques, D\'epartement de Physique, Facult\'e des Sciences,
Universit\'e Ferhat Abbas, S\'etif 19000, Algeria}

\affiliation{$^6$ Faculty of Physics, Lomonosov Moscow State
University, Moscow 119991, Russia}

\vskip 5mm

\date{\today}

\vskip 5mm

\begin{abstract}
We present kinematically complete theoretical calculations and experiments for transfer ionization in H$^++$He collisions at 630 keV/u. Experiment and theory are compared on the most detailed level of fully differential cross sections in the momentum space. This allows us to unambiguously identify contributions from the shake-off and binary encounter mechanisms of the reaction. It is shown that the simultaneous electron transfer and ionization is highly sensitive to the quality of a trial initial-state wave function.
\end{abstract}


\maketitle


\section{Introduction \label{introduction}}

Double ionization of a helium atom is the benchmark system to study electron-electron correlation in many electron systems and to test state-of-the-art theories. One \cite{Briggs2000jpb, Knapp2002prl},
two \cite{Kurka2009jpb} and multiphoton \cite{Doerner2002AP} double ionization has been explored in great detail experimentally and theoretically. For single photon absorption it has been shown that only fully differential cross sections reveal the mechanisms for double ionization as shake-off (SO), the so called two-step processes (TS) \cite{Knapp2002prl} and the quasi free mechanism (QFM) \cite{Amusia1975jpb,Galstyan2012pra}. For ion impact the state-of-the-art is much less satisfactory. Even for single ionization by fast particle impact, unresolved discrepancies between theory and experiment remain \cite{Schulz2003nature,Rescigno1999science, Foster2006prl}. This is even more problematic for transfer ionization (e. g. ${\rm H}^+ + {\rm He} \rightarrow {\rm H^0} + {\rm He}^{2+} + e^-$) \cite{Horsdal1986prl,Palinkas1989prl,Mergel1997prl,Mergel2001prl, Schmidt2002prl,Schmidtb2003epl,Schmidt2005pra}, where more interaction mechanisms than those of SO and TS, which are taken into account in the first Born approximation (FBA), contribute. So far no quantum theory has been able to calculate the full two dimensional momentum distributions. A theory which is
capable to predict all the experimental observed data for transfer ionization (TI) is a particular worthy goal as there are many indications in the literature that transfer ionization is an extremely interesting channel, whose rich features that cannot be accessed by photon, ion or electron impact double ionization or in strong laser pulses.

A physical explanation how a target electron can be captured into a bound state of a fast moving projectile within the single-interaction scenario was given by Oppenheimer, Brinkmann and Kramers (OBK) \cite{Oppenheimer1928pr,Jackson1953pr}. In the OBK approximation the electron transfer proceeds via a momentum space overlap of the initial target and final projectile wave function, which are displaced by the projectile velocity $v_p$. Thus kinematical capture at velocities above the Bohr velocity relies on the high-momentum components in the ground state wave function. Therefore the kinematical capture steeply decreases with projectile velocity ($\sigma \propto v_p^{-12}$) \cite{Oppenheimer1928pr,Belkic1978jpb}.

Emission of the second electron can take place via shake-off (SO) due the sudden removal of its correlation partner in the bound state \cite{Mcguire1988pra,Knapp2002prl,Godunov2004jpb,Schoeffler2005jpb}.
This suggests to treat the shake-off following a kinematical capture in analogy to shake-off following photoionization \cite{Mcguire1995jpb}, since in the OBK approximation as well as in the photoionization process the first electron is removed rapidly from a certain velocity component of the ground state (see \cite{Belkic1978jpb,Horsdal1982jpb,Mergel2001prl,Shi2002prl,Godunov2004jpb}).
Experiments of Mergel et al. \cite{Mergel1997prl,Mergel2001prl} raised the question whether higher angular momentum components in the ground-state wave function (so-called non s$^2$ contributions), though contributing only about 2~\% of the total wave function, play a significant role in producing the observed momentum distributions in the continuum. This has been later supported by calculations of Godunov and coworkers \cite{Shi2002prl,Godunov2004jpb,Schoeffler2005jpb,Godunov2005pra}.

Alternatively to shake-off also a second electron can be ejected as the result of direct collision with the projectile (so called  binary encounter, BE). This process is usually termed independent two-step-2 (TS2) mechanism \cite{Mcguire1977pra}. In its FBA version is presented by the schematic diagram $A_2$ in Fig. 1. Higher Born terms contribute to BE mechanism as well. After the first collision with a fast bare projectile a fast electron is ejected from the bound state. Subsequent (elastic) collisions with the nucleus in the intermediate state do not change its velocity too much. The intermediate $e-e$ interaction in contrast needs a more careful treatment.

Many approximations beyond the plane wave first Born approximation (PWFBA) exist. These are either in higher order of the Born series or approximate the higher order terms by using distorted waves. At high velocities the Thomas electron-nucleus and electron-electron mechanisms \cite{Thomas1927prsa,Briggs1979jpb,Mcguire1988pra,Mergel1997prl,Schmidt2005pra} become important. The classical description of this second order process, leading to capture of the electron, is given in the paper of Thomas \cite{Thomas1927prsa}. We note that in the classical physics, capture is considered to be a parallel motion of the proton and electron with the same velocity. To transfer an electron at highest velocities, a gamma-quant must be emitted to carry away the energy from the relative motion (so called radiative capture \cite{Eichler2007pr}). Instead of a gamma-quant the energy can be transferred to the remaining electron, which is emitted backward with respect to the projectile \cite{Voitkiv2008prl,Schulz2012prl}. Because of numerical difficulties the higher order processes are subject of future publications.

In this paper we present the experimental distribution of the momentum of the escaped electron in the scattering plane and the corresponding calculations in the PWFBA on the level of fully differential cross sections. This is the most sensitive test of the theory possible. The present calculations yield an unprecedented insight into the physical mechanisms at play as they allow to change the initial state and selectively switch on and off the distinct mechanisms discussed above. By comparing these calculations to our high-resolution experimental data, we separate ionization due to shake-off ($A_1+A_3$) or binary collision ($A_2$) (see Fig.~1) leading to distinct islands in momentum space. One of the results is that these data are extremely sensitive to the initial-state correlation.

\begin{figure}[htb]
  \begin{center}
   \epsfig{file=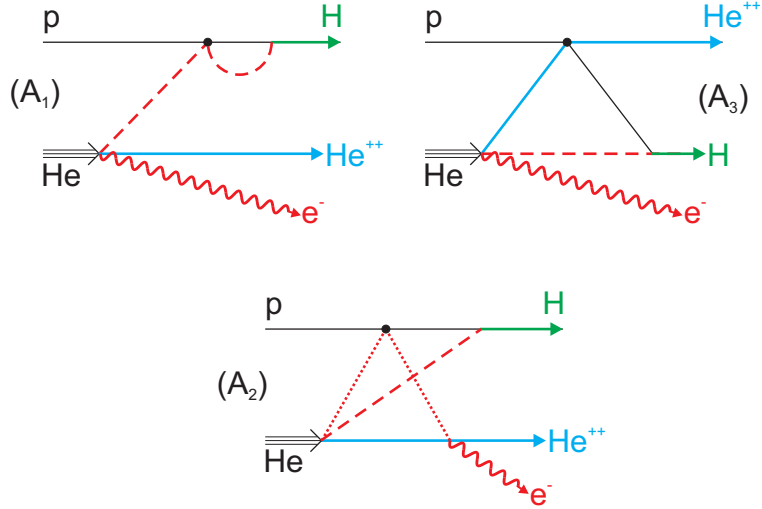,width=10cm}
    \caption{Schematic presentation of non-symmetrized $A_1+A_3$ and $A_2$ terms. $A_1$ (OBK) describes the collision between electron and proton followed by the capture of this electron by the projectile. The second electron is released due to rearrangement in the helium, known as shake-off (SO). The $A_3$ amplitude represents first an interaction between the projectile and the helium nucleus followed by electron capture. Similarly to $A_1$ the second electron is also released due to the sudden rearrangement in the helium. Term $A_2$ describes the classically termed TS2 amplitude. First the proton knocks-off a target electron into the continuum, followed by capturing the remaining electron from the helium.}
  \end{center}
  \label{fig0}
\end{figure}

In detail (see Fig. 1) the term $A_1$ (OBK) describes the collision between the proton and a target electron, which then is captured by the projectile. The second electron is released via SO. The collision between the proton and the target electron deflects the proton, leading to small scattering angles $<$0.5 mrad. Also in $A_3$ the electron emission takes place due SO, following the electron capture. In contrast to $A_1$ here, the capture proceeds after a nucleus-nucleus scattering between target and projectile followed by the electron capture. This term accesses also the larger scattering angles and dominates above 0.5 mrad.

In term $A_2$ the first interaction directly knocks-off a target electron and the second electron is captured. This TS2 process can be described in a first Born approximation. Strictly speaking the vertex H$^+ +$e$\to$H describes a bound state (respectively the electron transfer) and is not a subject of Born approximations. In principle, in the OBK term $A_1$ we can remove $e-H^+$ interaction and obtain the same matrix element. In this sense, the OBK term is the zeroth Born approximation.

Within OBK the electron is captured from the part of the initial bound state wave function, which has a high forward momentum component. In the case of strong correlation of electrons in the helium target, the electron ejected via shake-off should therefore be ejected preferentially in the backward direction, while in contrast the mechanism $A_2$ is responsible predominantly for the ejection of this electron in the forward direction. It was shown in \cite{Salim10} that the term $A_2$ can give a rather big and even the leading contribution to the differential cross section in the case of TI processes. Therefore we can expect to observe noticeable distributions both in forward and backward directions when the trial helium ground-state wave function is well correlated and in a multi configuration expansion contains higher angular momenta (non $s^2$ contributions).

\section{Experiment \label{experiment}}

We have used the COLTRIMS technique \cite{Ullrich1997jpb,Doerner2000pr,Ullrich2003rpp} to determine the momentum vectors of all final-state products. The experiment was performed at the Van de Graaff accelerator of the Institut f\"ur Kernphysik at the University of Frankfurt. The projectile beam (H$^+$) was collimated to a size of about 0.5 x 0.5 mm$^2$ at the target. 15 cm upstream of the target, a set of electrostatic deflector plates cleaned the primary beam from charge state impurities. The proton beam intersects with a supersonic helium gas jet (density of $5\times10^{11}$ atoms/cm$^2$ and a diameter of 1 mm). About 15 cm downstream a second set of electrostatic deflector plates separate the final charge state, thus only the neutral projectiles (H) hit a position and time sensitive multichannel plate (MCP) detector \cite{Jagutzki2002nima,Jagutzki2002nima2} yielding the projectile deflection angle and the time zero of the collision. The recoil ions were accelerated by a weak electrostatic field of 4.8 V/cm in the interaction region and detected on a 80 mm MCP-detector with delay-line anode. To optimize the resolution, a three dimensional time and space focussing geometry
\cite{Schoeffler2011njp,Mergel1995prl} was used for the spectrometer. A momentum resolution of 0.1 a.~u. was achieved in all three directions. The electrons were guided by a magnetic field of 15-25~Gauss and accelerated by the same electric field in a time focussing geometry \cite{Wiley1955rsi} onto a multi channel plate detector of 120 mm active diameter. A three-particle coincidence (H-He$^{2+}$-e) was applied to record the data event-by-event. From the positions of impact on the detectors and the time-of-flight we can derive the initial momentum vectors of the He$^{++}$ and the electron. Energy conservation was used for off-line background suppression. Furthermore the high resolution data allowed to distinguish data where the neutral projectile H$^0$ is found in an excited state from those, where the hydrogen is in its ground-state \cite{Kim2012pra}. Only these latter ones are presented in this letter.

\section{Theory \label{theory}}

Let us denote the projectile proton momentum by $\vec p_p$, the hydrogen momentum by $\vec p_H$, and the recoil-ion momentum by $\vec K$. We also define the transferred momentum as $\vec q=\vec p_H-\vec p_p$. We can deduce its approximate value using the momentum and energy conservation 

\begin{equation}
\vec q+\vec K+\vec k=0,
\end{equation}
\begin{equation}
\frac{p_p^2}{2m}+E_0^{He}=\frac{p_H^2}{2(m+1)}+\frac{K^2}{2M}+E^H+E^{ion}.
\end{equation}

Here $\vec k$ is the ejected electron momentum, the proton mass $m=1836.15$ , the helium ion mass $M\approx 4m$, $E_0^{He}\approx -2.903$, and $E^{ion}=k^2/2$.

Now we choose very small scattering angles for the outgoing hydrogen ($0\leq\theta_p\lesssim 0.5$ mrad). It leads to a practically zero ion velocity $K/M$ in the laboratory frame during the process, and we can consider the ion like immovable. The proton velocity $\vec v_p=\vec p/m$ varies about a few a.u. for its energy of several hundredths keV. This fact allows one to neglect ${K^2}/2M$ and $q^2/2m$ after insertion of $\vec p_H=\vec q+ \vec p_p$ into eq. (2). As a result we obtain 

\begin{equation}
\vec v_p\vec q=\frac12 v^2_p+Q; \quad Q=E_0^{He}-E^H-E^{ion},
\end{equation}

and choose the vector $\vec v_p$ as $z$-axis; there follows $q_z={v_p}/{2}+{Q}/{v_p}$\. The $x$-component of the vector is given by $\vec q$\ \ is $q_x=(\vec p_H)_x\approx mv_p\theta_p$.

In the presented experiments, the scattering plane $\{z,x,y=0\}$ formed by the momentum vectors $\vec p_p$ ($z$-axis) and $\vec p_H$ is fixed in space, and we put its polar angle $\phi=0$. The corresponding triple differential cross section (TDCS) takes the form

\begin{equation}
\frac{d^3\sigma}{dk_x
dk_zd\phi}=\frac{m^2}{(2\pi)^5}\int\limits_{\theta_i}^{\theta_{i+1}}\theta_pd\theta_p
\int\limits_{-\infty}^{\infty}dk_y |A_1+A_2+A_3|^2.
\end{equation}

Here $(\theta_i,\theta_{i+1})$ is the scattering angle domain and $(k_x,k_y,k_z)$ the electron momentum components. We calculate the TDCS depending on $(k_z,k_x)$ electron momentum distribution in
the scattering plane. We omit in short the mathematical and kinematical details of description of the symmetrized matrix elements $A_1$, $A_2$ and $A_3$, which are given in \cite{Salim10}.

In theoretical calculations we use two trial ground-state helium wave functions. One is the loosely correlated $1s^2$ Roothaan-Hartree-Fock (RHF) wave function \cite{RHF} (no angular correlation) with a rather poor ground-state energy of -2.861680 a. u. Another one is the highly correlated wave function given in \cite{Chuka06} with a ground-state energy of -2.903721 a. u. being very close to the experimental value of -2.903724377034 a. u.

\section{Results and Discussion \label{results}}

In Fig. 2 we present experimental electron momentum distributions and theoretical results in the scattering plane defined by the incoming projectile direction and the scattered projectile (the $x$-component of the vector $\vec p_H$ is positive here). Only events for a small projectile scattering angle $\theta_p\leq$0.25 mrad are selected. The experimental data in Fig. 2a show that at these small scattering angles, the electron is predominantly emitted in backward direction.

\begin{figure}[htb]
  \begin{center}
   \epsfig{file=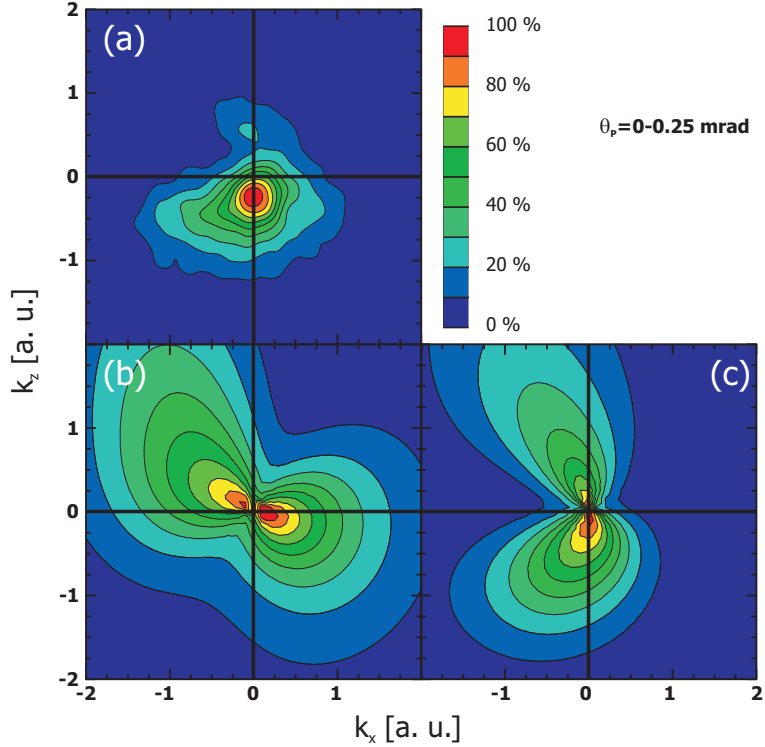,width=10cm}
    \caption{Experimental and theoretical data for 630 keV H+/He
    collisions for $\theta_p\leq 0.25$ mrad. (a) are the experimental data and the
    red dashed line represents the binary encounter ridge.
    (b) calculations using a helium $1s^2$ trial wave function,
    while (c) uses a highly correlated helium wave function with
    angular momentum, all including terms $A_1+A_2+A_3$}
  \end{center}
  \label{fig1}
\end{figure}

The results using a $1s^2$ trial helium wave function are shown in Fig. 2b. For small $\theta_p$ the momentum distribution is very similar to the binary- and recoil-peak structures (forward and backward) well known from electron impact ionization experiments \cite{Ehrhardt1986zp}. For comparison, the separation of the individual contributions/processes is shown in Fig. 3. The expected electron momentum distribution for the shake-off-process ($A_1+A_3$-term) in the case of loosely correlated helium wave function is shown in the top row (Fig. 3a), while in Fig. 3c (lower row) only the sequential TS2 mechanism ($A_2$-term) is
taken into account. The shake-off exhibits a perfectly isotropic behavior, as expected for a $1s^2$-state with zero angular momentum. This term has however a visible influence on the coherent sum of the different contributions $A_1$ + $A_2$ + $A_3$ (Fig. 2c) despite of a small overall dominance of slight dominance of the $A_2$ (TS2) term (the maximum in Fig. 3a is $5.75\times 10^{-7}$, while in Fig. 3c it is $9.00\times 10^{-7}$; the total maximum in Fig. 2b is $7.00\times 10^{-7}$). It changes the binary/recoil peak ratio, while conserving the general features of forward and backward contributions leaving the overall distributions to be similar. Comparing the experimental and theoretical results presented in Fig. 2a and Fig. 2b, we find noteable differences.

The agreement improves considerably for a well-correlated helium wave function with radial and angular $e-e$ correlations. In Fig. 2c the results of our calculations are shown again for small $\theta_p$ values. And they are split into the different
contributions in Fig. 3 (b,d). The maximum in Fig. 3b is $1.3\times 10^{-6}$, while in Fig. 3d it is $6.25\times 10^{-7}$;
the total maximum in Fig. 2c is $1.6\times 10^{-6}$. It can clearly be seen that the shake-off terms in Fig 3b show an asymmetric emission pattern (about 3 times larger compared to Fig. 3a), peaking in backward direction. A binary/recoil peak-like
structure is clearly visible again for the $A_2$-term using a correlated wave function (Fig. 3d). The coherent sum (Fig. 2c) also exhibits two clearly distinct non-equal peaks pointing forward and backward along the $z$-axis. This structure is considerably rotated clockwise compared to the one shown in Fig. 2b. Both calculations (Fig. 2c) and the experiment (Fig. 2a)
demonstrate predominantly the backward electron emission. However, detailed investigations of SO and TS2 contributions show that the term $A_2$ is still big and leads to an overestimate in the forward scattering domain.

\begin{figure}[htb]
  \begin{center}
    \epsfig{file=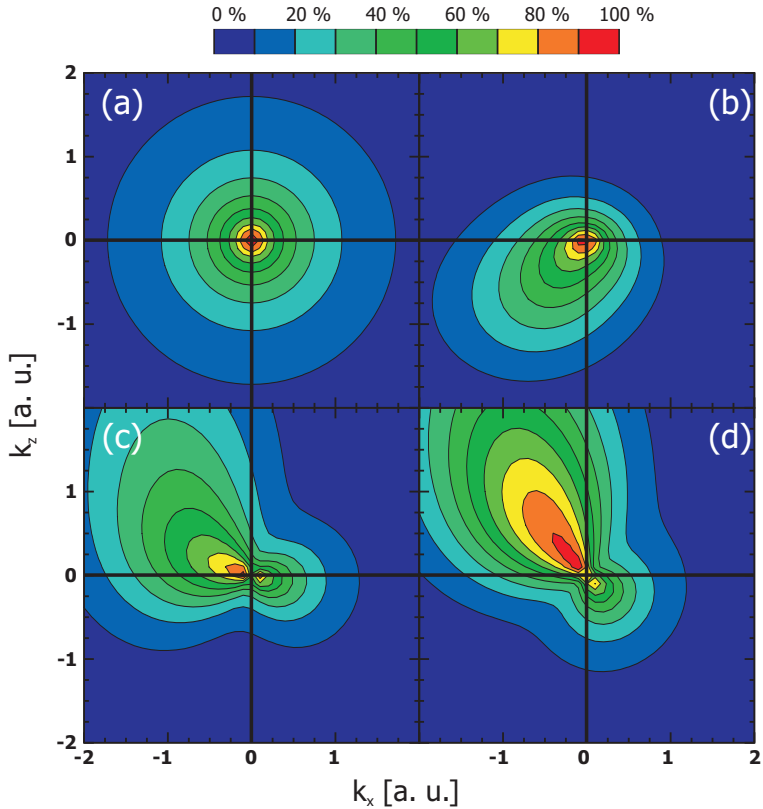,width=10cm}
    \caption{Calculations for $\theta_p\leq 0.25$ and helium $1s^2$ trial wave function separated in SO (a) and TS2 (c) contributions. Similar calculations for highly correlated helium wave function are presented in panels (b) and (d).}
  \end{center}
  \label{hylleraas}
\end{figure}

It is necessary to say a few words about contributions of second and higher order Born terms. After collision with the fast projectile proton the electron gets a rather high velocity and moves predominantly in the forward direction. It keeps this direction after elastic scattering on the atomic nucleus or another electron. We expect that the SO electrons are well described within FBA, whereas BE (TS2) electrons are more effected by higher (second) Born terms. As a consequence we can expect that the FBA term overestimates the contribution of forward scattered electrons (FBA and SBA matrix elements have different signs in total).

\begin{figure}[htb]
  \begin{center}
   \epsfig{file=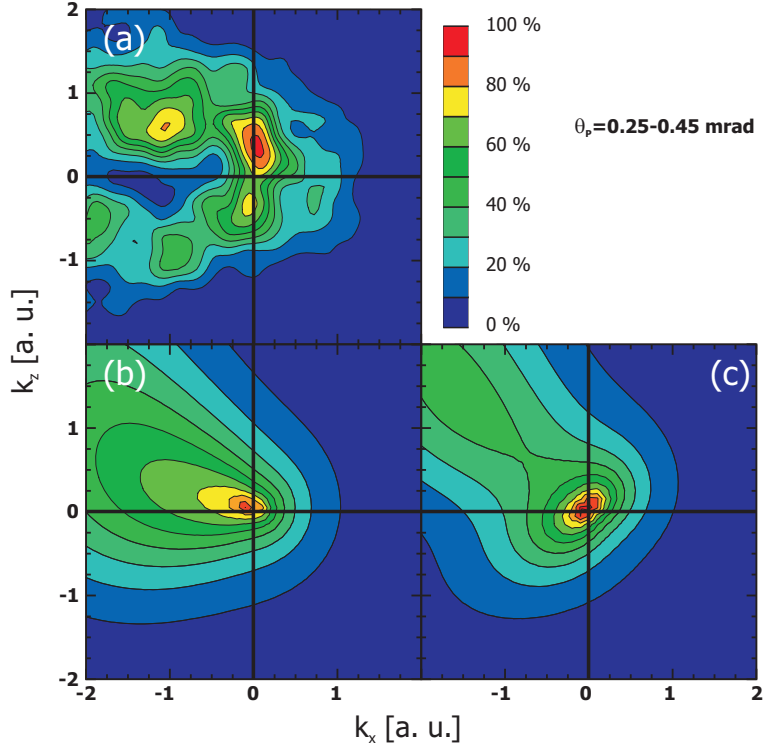,width=10cm}
    \caption{The same like in Fig. 2, but for $0.25\leq\theta_p\leq 0.45$ mrad.}
  \end{center}
  \label{correlated}
\end{figure}

We now consider plots corresponding to larger scattering angles 0.25$\leq\theta_p\leq$0.45 mrad (Fig. 4). The experiment (Fig. 4a) shows a richer of spots predominantly in forward direction and opposite to the $x$-component of $\vec p_H$. Now the FBA results (Fig. 4b for loosely correlated and Fig. 4c for highly correlated helium wave functions) are less structured. The correlated wave function displays some "pinch" structure at $\{k_x\sim -1, k_z\sim 0.4\}$, which we can be seen in Fig. 4a; but the main peak is well centered around $\{k_x=0, k_z=0\}$, while the experimental peak is notably shifted towards larger $k_z$. The predominant emission to the fourth quadrant is a result of rather hard binary collision which are selected in the plot by the projectile scattering angle.

We again can conclude that calculations with the correlated wave function give better agreement with the experiment, but now the limits of FBA clearly have been reached.

\section{Conclusions}

In conclusion, we presented highly differential theory (PWFBA) and experimental data from a kinematical complete experiment on transfer ionization in proton-helium collision at 630 keV/u. The observed splitting into forward and backward emission originates from two different contributions, the $A_2$-term (TS2, electron knock-off) and the $A_1+A_3$-term (shake-off). Comparison of a loosely and a strongly correlated wave function for the initial state confirms the high sensitivity of the experiment to the subtle features of the initial state wave function. FBA more or less explains the experiment at very small scattering angles and small electron momenta, but the SBA calculations are needed to improve results in forward scattering domain $k_z>0$. At bigger angles the SBA calculations are strongly needed.

\section{Acknowledgments}

This work was supported by Deutsche Forschungsgemeinschaft (DFG), grant SCHO 1210/2-1. As well this work partially supported by Russian Foundation of Basic Research (RFBR), grant 11-01-00523-a. All calculations were performed using Moscow State University Research Computing Centre (supercomputers Lomonosov and Chebyshev) and Joint Institute for Nuclear Research Central Information and Computer Complex. The authors are grateful to K. Kouzakov for useful discussions and help. We are also grateful for inspiring discussions with R. D\"orner and H. Schmidt-B\"ocking and continuous raised questions. We acknowledge helpful comments from A. Voitkiv.

\bibliographystyle{unsrt}

\end{document}